\documentclass[apjl]{emulateapj}

\usepackage{url}
\usepackage{enumerate}
\PassOptionsToPackage{hyphens}{url}\usepackage{hyperref}
\shorttitle{VHE detection of PKS~1441+25}
\shortauthors{Ahnen et al.}
\usepackage{scrextend}

\usepackage{color}
\usepackage[normalem]{ulem}

\begin{document}

\dimen\footins=2\baselineskip\relax

\title{Very-high-energy $\gamma$ rays from the Universe's middle age:\\ detection of the \lowercase{$z$}=0.940 blazar PKS~1441+25 with MAGIC}

\author{
  M.~L.~Ahnen\altaffilmark{1},
  S.~Ansoldi\altaffilmark{2},
  L.~A.~Antonelli\altaffilmark{3},
  P.~Antoranz\altaffilmark{4},
  A.~Babic\altaffilmark{5},
  B.~Banerjee\altaffilmark{6},
  P.~Bangale\altaffilmark{7},
  U.~Barres de Almeida\altaffilmark{7,26},
  J.~A.~Barrio\altaffilmark{8},
  W.~Bednarek\altaffilmark{9}
% ...
}

\altaffiltext{$\dagger$} {\url{Josefa.Becerra@nasa.gov}}
\altaffiltext{$\ddagger$} {\url{mnievas@ucm.es}}
\altaffiltext{$\sharp$} {\url{manganaro@iac.es}}
\altaffiltext{$\S$} {\url{fabrizio.tavecchio@brera.inaf.it}}

\dimen\footins=200\baselineskip\relax

\altaffiltext{1} {ETH Zurich, CH-8093 Zurich, Switzerland}
\altaffiltext{2} {Universit\`a di Udine, and INFN Trieste, I-33100 Udine, Italy}
\altaffiltext{3} {INAF National Institute for Astrophysics, I-00136 Rome, Italy}
\altaffiltext{4} {Universit\`a  di Siena, and INFN Pisa, I-53100 Siena, Italy}
\altaffiltext{5} {Croatian MAGIC Consortium, Rudjer Boskovic Institute, University of Rijeka, University of Split and University of Zagreb, Croatia}
\altaffiltext{6} {Saha Institute of Nuclear Physics, 1\textbackslash{}AF Bidhannagar, Salt Lake, Sector-1, Kolkata 700064, India}
\altaffiltext{7} {Max-Planck-Institut f\"ur Physik, D-80805 M\"unchen, Germany}
\altaffiltext{8} {Universidad Complutense, E-28040 Madrid, Spain}
\altaffiltext{9} {University of \L\'od\'z, PL-90236 Lodz, Poland}
\altaffiltext{10} {Deutsches Elektronen-Synchrotron (DESY), D-15738 Zeuthen, Germany}
\altaffiltext{11} {IFAE, Campus UAB, E-08193 Bellaterra, Spain}
\altaffiltext{12} {Universit\"at W\"urzburg, D-97074 W\"urzburg, Germany}
\altaffiltext{13} {Centro de Investigaciones Energ\'eticas, Medioambientales y Tecnol\'ogicas, E-28040 Madrid, Spain}
\altaffiltext{14} {Inst. de Astrof\'isica de Canarias, E-38200 La Laguna, Tenerife, Spain; Universidad de La Laguna, Dpto. Astrof\'isica, E-38206 La Laguna, Tenerife, Spain}
\altaffiltext{15} {Universit\`a di Padova and INFN, I-35131 Padova, Italy}
\altaffiltext{16} {Institute for Space Sciences (CSIC\textbackslash{}IEEC), E-08193 Barcelona, Spain}
\altaffiltext{17} {Technische Universit\"at Dortmund, D-44221 Dortmund, Germany}
\altaffiltext{18} {Unitat de F\'isica de les Radiacions, Departament de F\'isica, and CERES-IEEC, Universitat Aut\`onoma de Barcelona, E-08193 Bellaterra, Spain}
\altaffiltext{19} {Universitat de Barcelona, ICC, IEEC-UB, E-08028 Barcelona, Spain}
\altaffiltext{20} {Japanese MAGIC Consortium, ICRR, The University of Tokyo, Department of Physics and Hakubi Center, Kyoto University, Tokai University, The University of Tokushima, KEK, Japan}
\altaffiltext{21} {Finnish MAGIC Consortium, Tuorla Observatory, University of Turku and Department of Physics, University of Oulu, Finland}
\altaffiltext{22} {Inst. for Nucl. Research and Nucl. Energy, BG-1784 Sofia, Bulgaria}
\altaffiltext{23} {Universit\`a di Pisa, and INFN Pisa, I-56126 Pisa, Italy}
\altaffiltext{24} {ICREA and Institute for Space Sciences (CSIC\textbackslash{}IEEC), E-08193 Barcelona, Spain}
\altaffiltext{25} {Universit\`a dell'Insubria and INFN Milano Bicocca, Como, I-22100 Como, Italy}
\altaffiltext{26} {now at Centro Brasileiro de Pesquisas F\'isicas (CBPF\textbackslash{}MCTI), R. Dr. Xavier Sigaud, 150 - Urca, Rio de Janeiro - RJ, 22290-180, Brazil}
\altaffiltext{27} {Humboldt University of Berlin, Istitut f\"ur Physik  Newtonstr. 15, 12489 Berlin Germany}
\altaffiltext{28} {now at Ecole polytechnique f\'ed\'erale de Lausanne (EPFL), Lausanne, Switzerland}
\altaffiltext{29} {Istituto Nazionale di Fisica Nucleare, Sezione di Padova, I-35131 Padova, Italy}
\altaffiltext{30} {now at Laboratoire AIM, Service d'Astrophysique, DSM\textbackslash{}IRFU, CEA\textbackslash{}Saclay FR-91191 Gif-sur-Yvette Cedex, France}
\altaffiltext{31} {now at Finnish Centre for Astronomy with ESO (FINCA), Turku, Finland}
\altaffiltext{32} {also at INAF-Trieste}
\altaffiltext{33} {Istituto Nazionale di Fisica Nucleare, Sezione di Trieste, I-34127 Trieste, Italy}
\altaffiltext{34} {also at ISDC - Science Data Center for Astrophysics, 1290, Versoix (Geneva)}
\altaffiltext{35} {Department of Physics and Astronomy, Clemson University, Kinard Lab of Physics, Clemson, SC 29634-0978, USA}
\altaffiltext{36} {Universit\`a di Pisa and Istituto Nazionale di Fisica Nucleare, Sezione di Pisa I-56127 Pisa, Italy}
\altaffiltext{37} {W. W. Hansen Experimental Physics Laboratory, Kavli Institute for Particle Astrophysics and Cosmology, Department of Physics and SLAC National Accelerator Laboratory, Stanford University, Stanford, CA 94305, USA}
\altaffiltext{38} {Dipartimento di Fisica, Universit\`a di Trieste, I-34127 Trieste, Italy}
\altaffiltext{39} {Dipartimento di Fisica e Astronomia ``G. Galilei'', Universit\`a di Padova, I-35131 Padova, Italy}
\altaffiltext{40} {NASA Goddard Space Flight Center, Greenbelt, MD 20771, USA}
\altaffiltext{41} {Department of Physics and Department of Astronomy, University of Maryland, College Park, MD 20742, USA}
\altaffiltext{42} {Istituto Nazionale di Fisica Nucleare, Sezione di Pisa, I-56127 Pisa, Italy}
\altaffiltext{43} {Istituto Nazionale di Fisica Nucleare, Sezione di Bari, I-70126 Bari, Italy}
\altaffiltext{44} {Istituto Nazionale di Fisica Nucleare, Sezione di Torino, I-10125 Torino, Italy}
\altaffiltext{45} {Dipartimento di Fisica Generale ``Amadeo Avogadro'' , Universit\`a degli Studi di Torino, I-10125 Torino, Italy}
\altaffiltext{46} {Laboratoire Univers et Particules de Montpellier, Universit\'e Montpellier, CNRS/IN2P3, Montpellier, France}
\altaffiltext{47} {Laboratoire Leprince-Ringuet, \'Ecole polytechnique, CNRS/IN2P3, Palaiseau, France}
\altaffiltext{48} {Consorzio Interuniversitario per la Fisica Spaziale (CIFS), I-10133 Torino, Italy}
\altaffiltext{49} {INAF-Istituto di Astrofisica Spaziale e Fisica Cosmica, I-20133 Milano, Italy}
\altaffiltext{50} {Agenzia Spaziale Italiana (ASI) Science Data Center, I-00133 Roma, Italy}
\altaffiltext{51} {Istituto Nazionale di Fisica Nucleare, Sezione di Perugia, I-06123 Perugia, Italy}
\altaffiltext{52} {INAF Osservatorio Astronomico di Roma, I-00040 Monte Porzio Catone (Roma), Italy}
\altaffiltext{53} {INAF Istituto di Radioastronomia, I-40129 Bologna, Italy}
\altaffiltext{54} {Dipartimento di Astronomia, Universit\`a di Bologna, I-40127 Bologna, Italy}
\altaffiltext{55} {Universit\`a Telematica Pegaso, Piazza Trieste e Trento, 48, I-80132 Napoli, Italy}
\altaffiltext{56} {Universit\`a di Udine, I-33100 Udine, Italy}
\altaffiltext{57} {Dipartimento di Fisica ``M. Merlin'' dell'Universit\`a e del Politecnico di Bari, I-70126 Bari, Italy}
\altaffiltext{58} {Laboratoire AIM, CEA-IRFU/CNRS/Universit\'e Paris Diderot, Service d'Astrophysique, CEA Saclay, F-91191 Gif sur Yvette, France}
\altaffiltext{59} {NASA Postdoctoral Program Fellow, USA}
\altaffiltext{60} {University of North Florida, Department of Physics, 1 UNF Drive, Jacksonville, FL 32224 , USA}
\altaffiltext{61} {Department of Physics, KTH Royal Institute of Technology, AlbaNova, SE-106 91 Stockholm, Sweden}
\altaffiltext{62} {The Oskar Klein Centre for Cosmoparticle Physics, AlbaNova, SE-106 91 Stockholm, Sweden}
\altaffiltext{63} {Space Science Division, Naval Research Laboratory, Washington, DC 20375-5352, USA}
\altaffiltext{64} {Dipartimento di Fisica, Universit\`a degli Studi di Perugia, I-06123 Perugia, Italy}
\altaffiltext{65} {Catholic University of America, Washington, DC 20064, USA}
\altaffiltext{66} {University of Maryland, Baltimore County, Baltimore, MD 21250, USA} 
\altaffiltext{67} {Hiroshima Astrophysical Science Center, Hiroshima University, Higashi-Hiroshima, Hiroshima 739-8526, Japan}
\altaffiltext{68} {Istituto Nazionale di Fisica Nucleare, Sezione di Roma ``Tor Vergata'', I-00133 Roma, Italy}
\altaffiltext{69} {Funded by contract FIRB-2012-RBFR12PM1F from the Italian Ministry of Education, University and Research (MIUR)}
\altaffiltext{70} {Institut f\"ur Astro- und Teilchenphysik and Institut f\"ur Theoretische Physik, Leopold-Franzens-Universit\"at Innsbruck, A-6020 Innsbruck, Austria}
\altaffiltext{71} {NYCB Real-Time Computing Inc., Lattingtown, NY 11560-1025, USA}
\altaffiltext{72} {Solar-Terrestrial Environment Laboratory, Nagoya University, Nagoya 464-8601, Japan}
\altaffiltext{73} {Department of Physical Sciences, Hiroshima University, Higashi-Hiroshima, Hiroshima 739-8526, Japan}
\altaffiltext{74} {Cahill Center for Astrophysics, 1216 East California Boulevard, California Institute of Technology, Pasadena, CA 91125, USA} 
\altaffiltext{75} {Tuorla Observatory, Department of Physics and Astronomy, University of Turku, Finland}
\altaffiltext{76} {Instituto Nacional de Astrof\'isica \'Optica y Electr\'onica (INAOE), Apartado Postal 51 y 216, 72000 Puebla, M\'exico} 
\altaffiltext{77} {Naturwissenschaftliches Labor f\"ur Sch\"uler, Friedrich-Koenig-Gymnasium, W\"urzburg}
\altaffiltext{78} {Aalto University Mets\"ahovi Radio Observatory, Mets\"ahovintie 114, 02540 Kylm\"al\"a, Finland}
\altaffiltext{79} {INAF-Istituto di Astrofisica e Planetologia Spaziale, Via Fosso del Cavaliere 100, I-00133 Rome, Italy} 
\altaffiltext{80} {Aalto University Department of Radio Science and Engineering,P.O. BOX 13000, FI-00076 AALTO, Finland.}


\begin{abstract}
The flat-spectrum radio quasar PKS~1441+25 at a redshift of $z=0.940$ is
detected between 40 and 250\,GeV with a significance of 25.5\,$\sigma$
using the MAGIC telescopes. Together with the gravitationally lensed blazar 
QSO~B0218+357 ($z = 0.944$), PKS~1441+25 is the most distant very high energy (VHE) blazar detected to date. 
The observations were triggered by an outburst in 2015 April seen at GeV energies 
with the Large Area Telescope on board {\it Fermi}. Multi-wavelength observations
suggest a subdivision of the high state into two distinct flux states.
In the band covered by MAGIC, the variability time scale is estimated to be $ 6.4 \pm 1.9$ days. 
Modeling the broadband spectral energy distribution with an external Compton model, 
the location of the emitting region is understood as originating in the jet 
outside the broad line region (BLR) during the period of high activity, while 
 being partially within the BLR during the period of low 
(typical) activity. 
The observed VHE spectrum during the highest activity is used to probe the extragalactic background light at an unprecedented 
distance scale for ground-based gamma-ray astronomy.
\end{abstract}


\section{Introduction}
PKS~1441+25 is a known high-energy (HE, $0.1 \,\rm{GeV}<E<100\,\rm{GeV}$) $\gamma$-ray flat spectrum radio quasar (FSRQ) \citep{1FGL,Nolan12,1FHL} located at $z=0.9397\pm0.0003_{stat}$\footnote{From SDSS: \url{http://skyserver.sdss.org/dr10/en/get/SpecById.ashx?id=6780257851631206400}, see also \cite{shaw12}}. In January 2015 it became active from $\gamma$ rays to the near-infrared \citep{atel6878,atel6895,atel6923,atel7402}. In April, the detection of the source with a hard spectral index with the \textit{Fermi Gamma-ray Space} Large Area Telescope (LAT) together with increased multi-wavelength (MWL) emission triggered the MAGIC observations. They resulted in the first detection of this source at very high energy (VHE, $E>100$~GeV) \citep[][]{atel7416}, later followed up by VERITAS \citep{atel7433}. This detection makes PKS1441+25 the\footnote{\url{http://tevcat.uchicago.edu}} 5th FSRQ with a firm classification detected at VHE, and the most distant known VHE source, along with QSO B0218+357 \citep[$z=0.944\pm0.002$,][]{SitarekICRC2015}.

In this letter, the MWL observations are discussed in the context of an external Compton model for four different states of activity, dubbed periods A (MJD 57125.0--57130.0), B (57130.0--57135.5), C (57135.5--57139.5) and D (57149.0--57156.0). Upper limits on the extragalactic background light (EBL) are obtained in the VHE band.

\section{Observations and analysis}

\subsection{VHE $\gamma$-ray observations}

MAGIC is a stereoscopic system consisting of two 17\,m diameter Imaging Atmospheric Cherenkov Telescopes located at the Observatorio del Roque de los Muchachos, on the Canary Island of La Palma. The current sensitivity for low-zenith observations ($zd<30^\circ$) above 220\,GeV is $0.66\pm0.03\,\%$ of the Crab Nebula's flux in 50~hr \citep{Aleksic2015_Upgrade2}.

The MAGIC telescopes monitored PKS~1441+25 from 2015 April 18 to 27 (MJD 57130--57139, for a total of 29.9~hr) and May 8-9 (MJD 57150--57151, for 1.8~hr), the observational gap being due to the full-moon break. The observations were performed in wobble mode with a $0^{\circ}.4$ offset and four symmetric positions with respect to the camera center \citep{fomin1994}. The data were collected in the zenith angle range of $3^\circ<zd<38^\circ$.

The analysis of the data is performed using the standard MAGIC analysis framework \emph{MARS} \citep{Zanin2013_MARS, Aleksic2015_Upgrade2} and Monte Carlo (MC) simulations matching the night-sky background levels. 

PKS~1441+25 is detected with a significance of $25.5\,\sigma$ ($\gamma$-ray like excess $N_{ex}=4010\pm160$) during periods \emph{B}+\emph{C}. No significant emission was found in period \emph{D}.

The differential VHE spectrum is measured from 40 to 250\,GeV and 50 to 160\,GeV in periods \emph{B} and \emph{C} respectively. A power-law (PWL) can describe both observed and EBL-corrected spectra using the model of \cite{Dominguez2011}:

\begin{center}
  \begin{equation}
    \frac{dF}{dE}=f_0 \left(\frac{E}{100\,\mathrm{GeV}}\right)^{-\Gamma},
    \label{observed_spectrum}
  \end{equation}
\end{center}
where the normalization constant $f_0$, the spectral index $\Gamma$ and the goodness of the fit ($\chi^2/ndf$ and $p$-value) are:

\begin{enumerate}
\item Period \emph{B}:
  \begin{enumerate}[(i)]
  \item Observed: $f_{0}=(1.14\pm0.06_{stat}\pm0.20_{sys}) \times 10^{-9}$ $\mathrm{cm}^{-2} \mathrm{s}^{-1} \mathrm{TeV}^{-1}$ and $\Gamma = 4.62\pm0.11_{stat}\pm0.18_{sys}$ ($\chi^2/ndf=22.9/6$, $P=8.4\times 10^{-4}$)
  \item EBL-corrected: $f_{0}=(2.7\pm0.1_{stat}\pm0.5_{sys}) \times 10^{-9}$ $\mathrm{cm}^{-2} \mathrm{s}^{-1} \mathrm{TeV}^{-1}$ and $\Gamma =3.18\pm0.15_{stat}\pm0.18_{sys}$ ($\chi^2/ndf=5.6/6$, $P=0.47$)
\end{enumerate}
\item Period \emph{C}:
  \begin{enumerate}[(i)]
  \item Observed: $f_{0}=(0.82\pm0.09_{stat}\pm0.13_{sys}) \times 10^{-9}$ $\mathrm{cm}^{-2} \mathrm{s}^{-1} \mathrm{TeV}^{-1}$ and $\Gamma =3.7\pm0.4_{stat}\pm0.2_{sys}$ ($\chi^2/ndf=2.7/3$, $P=0.44$)
  \item EBL-corrected: $f_{0}=(1.7\pm0.2_{stat}\pm0.3_{sys}) \times 10^{-9} \mathrm{cm}^{-2} \mathrm{s}^{-1} \mathrm{TeV}^{-1}$ and $\Gamma =2.5\pm0.4_{stat}\pm0.2_{sys}$ ($\chi^2/ndf=4.3/3$, $P=0.23$)
\end{enumerate}
\end{enumerate}

From a likelihood ratio test (LRT), a model with intrinsic curvature such as a log-parabola (LP) is preferred at $4.2\,\sigma$ during period \emph{B}. It is defined as:

\begin{center}
  \begin{equation}
    \frac{dF}{dE}=f_0 \left(\frac{E}{100\,\mathrm{GeV}}\right)^{-\Gamma_{LP}-b\log_{10}{ \frac{E}{100\,\mathrm{GeV}}}},
    \label{LP}
  \end{equation}
\end{center}
where $f_{0}=(1.39\pm0.09_{stat}\pm0.24_{sys}) \times 10^{-9} \mathrm{cm}^{-2} \mathrm{s}^{-1}$ $\mathrm{TeV}^{-1}$, $\Gamma_{LP} =4.69\pm0.16_{stat}$ and $b=3.2\pm1.0_{stat}$ ($\chi^2/ndf=5.2/5$, $P=0.39$). A full description of the MAGIC systematic uncertainties can be found in \cite{Aleksic2015_Upgrade2} and references therein.


\subsection{HE $\gamma$-ray observations}

In nominal survey mode the LAT monitors the entire $\gamma$-ray sky every 3~hr in the energy range from $20\,\mathrm{MeV}$ to at least $300\,\mathrm{GeV}$ \citep{Atwood09}. We select Pass 8 SOURCE class events collected from 2015 April 8 to May 23 (MJD 57120--57165) between 100 MeV to 500 GeV and within a $10^\circ$ Region of Interest (ROI) centered at the location of PKS~1441+25. In order to reduce contamination from the Earth Limb, a zenith angle cut of $<90^\circ$ is applied. The analysis is performed with the ScienceTools software package version v10r0p5 using the \texttt{P8R2\_SOURCE\_V6}\footnote{\url{http://fermi.gsfc.nasa.gov/ssc/data/analysis/documentation/Cicerone/Cicerone_LAT_IRFs/IRF_overview.html}} instrument response functions and the \texttt{gll\_iem\_v06} and \texttt{iso\_P8R2\_SOURCE\_V6\_v06} models\footnote{\url{http://fermi.gsfc.nasa.gov/ssc/data/access/lat/BackgroundModels.html}} for the Galactic and isotropic diffuse emission, respectively.

The likelihood fit is performed using \texttt{gtlike}, including all 3FGL sources \citep{Acero15} within $20^\circ$ from PKS~1441+25. The spectral indices and fluxes are left free for sources within $10^\circ$, while sources from $10^\circ$ to $20^\circ$ have their parameters fixed to the catalog value. Both the flux and the spectral index of PKS~1441+25 are left free for the light curve calculation, while the parameters for the rest of the sources in the ROI are fixed except the diffuse components. Five photons of energies 10-50\,GeV were detected with a probability of association with PKS~1441+25 larger than $99.6\,\%$, calculated with \texttt{gtsrcprob}. The spectrum of PKS~1441+25 is well fit by a PWL (as in the 3FGL catalog) and no significant curvature was found. During the flare (period B+C), the spectral index is $\Gamma=1.75\pm0.06$, harder than the 3FGL value $\Gamma_{3FGL}=2.13\pm0.07$.


\subsection{Hard X-ray observations}

{\it NuSTAR} (Nuclear Spectroscopic Telescope Array; \citealt{harrison+2013}) is a hard X-ray telescope operating in the energy range between $3$ and 79\,keV. PKS~1441+25 was observed with {\it NuSTAR} on 2015 April~25--26 (MJD~57137.1113) for a total (on-source) exposure of 40~ks. The data are processed using the standard \texttt{nupipeline} script (version 1.4.1) available in the NuSTARDAS software package \citep{perri+2014}. The source spectrum extends up to $\simeq$25~keV, and can be described by a PWL with spectral index $\Gamma=2.30\pm0.08$ ($\chi^2/ndf=10.4/7$). No significant variability is detected during the observation.


\subsection{X-ray and optical--UV observations}

A \textit{Swift} \citep{Gehrels2004} target of opportunity started on 2015 April 15. \textit{Swift}-XRT \citep{Burrows2005} observed the source in photon-counting mode. Standard filtering and analysis of the data were employed. The source exhibited a soft X-ray photon index (from $1.94 \pm 0.16$ to $2.55 \pm 0.24$) and is described by an absorbed PWL model, with the Galactic absorption fixed to $N_H = 3.2 \times$ $10^{20} \mathrm{cm}^{-2}$ \citep{kalberla2005} during April--May 2015. For comparison, the observations on 2010 June 12 (MJD 55359) can be fit with a PWL with spectral index $1.2\pm0.3$

The \textit{Swift}-UVOT \citep{Roming2005} flux in several bands was estimated using aperture photometry. De-reddening is performed using $E(B-V)=0.033$ \citep{Schlafly2011} and $A_V/E(B-V)=3.1$ \citep{Schultz1975}.


\subsection{Optical observations}

Optical R-band observations started on MJD 57130 and were performed using the 35\,cm Celestron telescope attached to the KVA\footnote{\url{http://users.utu.fi/kani/1m}} 60\,cm telescope (La Palma, Canary Islands, Spain) and the 50 cm Hans-Haffner-Telescope (Hettstadt, W\"urzburg, Germany).\footnote{\url{http://schuelerlabor-wuerzburg.de/?p=Sternwarte}} The data are analyzed using differential photometry and corrected for Galactic extinction \citep{Schlafly2011}. The host galaxy contribution is negligible compared to the flux of the source during these observations. The optical emission shows a high degree of polarization, reaching a maximum of $37.7\,\%$ on MJD~57132 \citep{Smith2015}.


\subsection{Near infrared observations}
NIR observations in the J, H, and K$_{\mathrm{S}}$ bands started on MJD 57141 with CANICA,\footnote{\url{http://www.inaoep.mx/~astrofi/cananea/canica/}} a direct camera at the 2.1\,m telescope of the Guillermo Haro Observatory located at Cananea, Mexico. The flux is estimated by means of differential photometry using the 2MASS catalog \citep{Skrutskie2006}.

\subsection{Radio observations}

The observations of PKS~1441+25 with the Mets\"ahovi 13.7-m radio telescope started on MJD 57135. The measurements were made with a 1 GHz-band dual beam receiver centered at 37 GHz. A detailed description of the observation and analysis methods can be found in \cite{Terasranta}.

\section{Results}\label{Section:Results}


\subsection{\emph{Multi-wavelength} flux evolution}

\begin{figure*}
  \centering
  \includegraphics[width=0.75\textwidth]{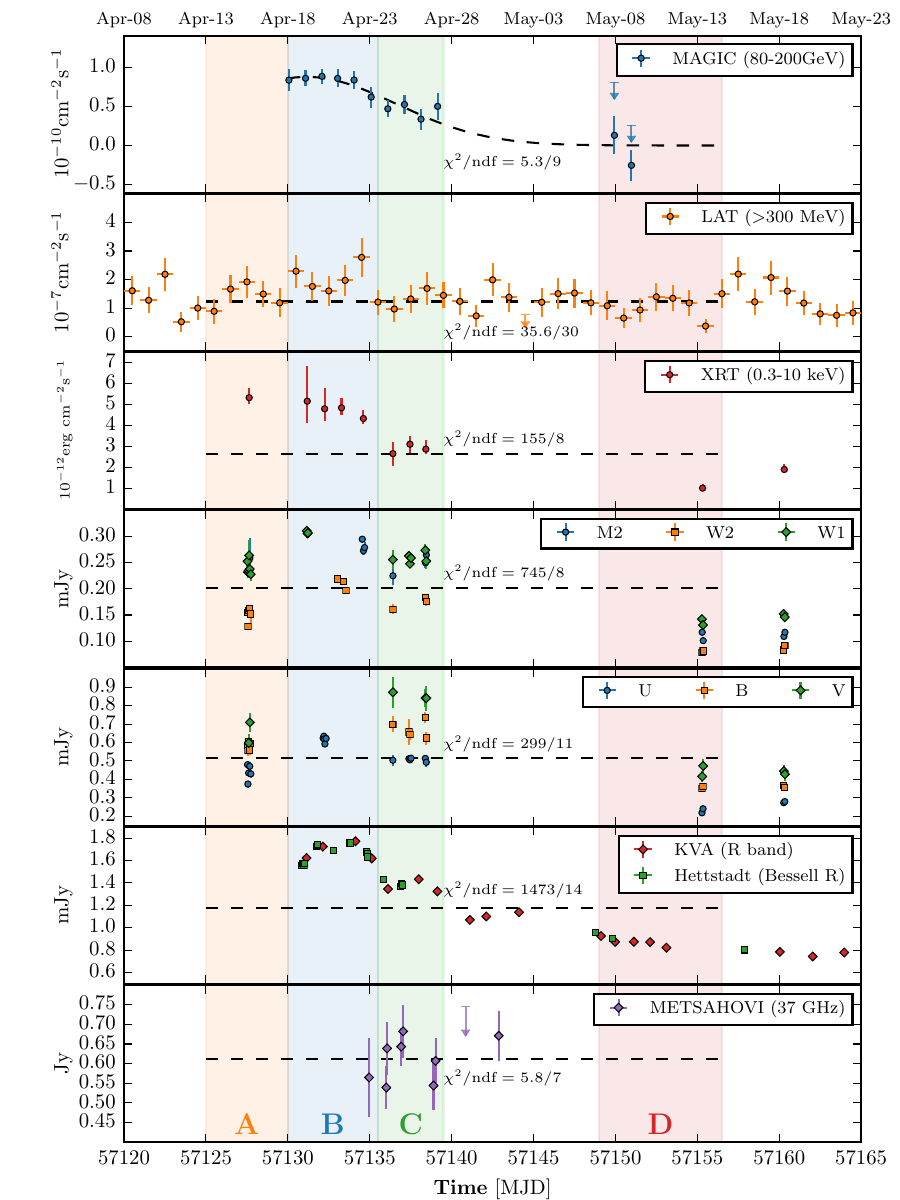}
  \caption{Light curve of PKS~1441+25 at different wavelengths. The shaded areas marked as A, B, C, D depict the different states of the source considered in Sec.~\ref{BroadbandSED}. Only filters ``UVOT-M2'', ``UVOT-B'' and ``KVA-R'' are used in the fit in the optical--UV bands.}
  \label{fig:MWL_LC}
\end{figure*}


The MWL light curve is presented in Fig.~\ref{fig:MWL_LC}. In the VHE band, the no-variability hypothesis can be discarded as it results in a $\chi^2/ndf=52.5/11$ ($P^{B-D}_{const}=2.2\times10^{-7}$) for \emph{B}+\emph{C}+\emph{D}. A constant fit is also unlikely for the flare in April (\emph{B}+\emph{C}) with a $\chi^2/ndf=26.0/9$ ($P^{B+C}_{const}=2.1\times10^{-3}$). We gauge the characteristic variability time scale by heuristically fitting the VHE light curve with a Gaussian function. The fit provides a standard deviation $\sigma=5.5 \pm 1.6\,\mathrm{days}$ (halving flux time of $6.4\pm1.9\,\mathrm{days}$) and a peak flux of $(8.8\pm0.6) \times 10^{-11}\, \mathrm{cm}^{-2}\, \mathrm{s}^{-1}$ ($\chi^2/ndf=5.3/9$, $P^{B-D}_{Gaussian}=0.80$). For X-rays, a halving flux time of $7.6\pm1.7\,\mathrm{days}$ was found.

The average flux in \emph{B} is larger than in \emph{C} by a factor of $F_B/F_C = 1.80 \pm 0.27$ in VHE. A similar pattern was found in X-rays ($F_B/F_C=1.58 \pm 0.17$), optical ($F_B/F_C=1.23 \pm 0.02$) and a hint in the HE ($1.40 \pm 0.29$). No intra-night variability is detected.

\subsection{Broadband spectral energy distribution\label{BroadbandSED}}

\begin{figure*}
  \centering
  \includegraphics[width=0.68\textwidth]{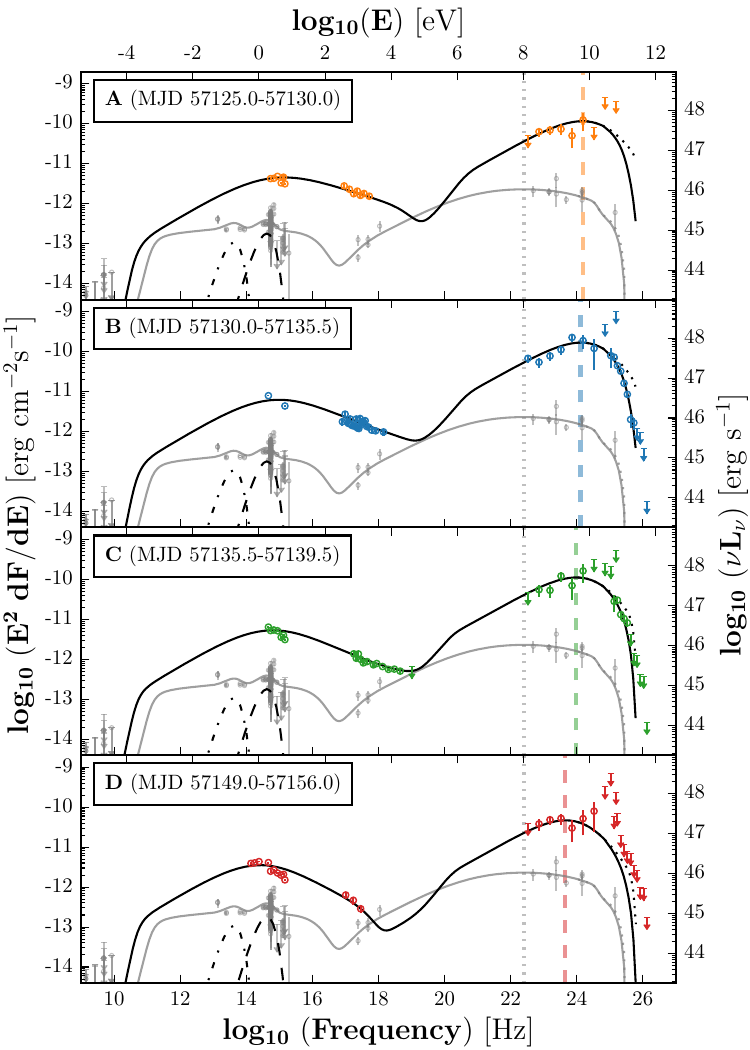}
  \caption{MWL SEDs for PKS~1441+25 for the four states of the source indicated in Fig.~\ref{fig:MWL_LC}. The broadband emission model for the observed (solid line) and EBL-de-absorbed (dotted line) spectrum, using {the model of} \cite{Dominguez2011}, together with the disk (dashed) and torus (dash--dotted) emission component are shown. Archival data extracted from ASDC (\protect\url{http://tools.asdc.asi.it}) are shown in gray. The VHE spectral points are not corrected for EBL absorption. Vertical lines indicate the IC peaks.}
  \label{fig:MWL_SED}
\end{figure*}

\begin{deluxetable*}{cccccccccc}
\tabletypesize{\small}
\tablewidth{\textwidth}
\tablecaption{Input parameters for the emission models shown in Fig.~\ref{fig:MWL_SED}}
\tablenum{1}

\tablehead{\colhead{Period} & \colhead{MJD} & \colhead{$\gamma_{\rm min}$} & \colhead{$\gamma_{\rm b}(10^4)$} & \colhead{$\gamma_{\rm max}(10^6)$} & \colhead{$n_2$} & \colhead{$B$ (G)} & \colhead{$K$ ($10^{3}\,\mathrm{cm}^{-3}$)} & \colhead{$\nu_{IC}[Hz]$} & \colhead{CD} \\}

\startdata
\emph{A}  &  57125.0--57130.0 &  $80$  &  $1.0$ & $1.0$  & $3.55$  &  $0.15$  &  $2.80$ & $24.2$ & $24$ \\
\emph{B}  &  57130.0--57135.5 &  $80$  &  $1.0$ & $1.0$  & $3.70$  &  $0.15$  &  $4.00$ & $24.1$ & $25$ \\
\emph{C}  &  57135.5--57139.5 &  $50$  &  $0.8$ & $1.0$  & $3.75$  &  $0.17$  &  $3.35$ & $24.0$ & $21$ \\
\emph{D}  &  57149.0--57156.0 &  $50$  &  $0.5$ & $0.2$  & $3.90$  &  $0.23$  &  $2.00$ & $23.6$ & $13$ \\
Archival &  -          &  $20$  &  $10^{-2}$  & $3\times10^{-2}$ &  $3.05$ & $0.35$ & $70$ & $ 22.4$ &  $7$ \\
\enddata

\tablecomments{The other parameters are kept fixed (see text). The IC peak frequency (in logarithmic scale) and the Compton Dominance (CD) are also reported.}
\label{tab:model}
\end{deluxetable*}

The MWL spectral energy distributions (SEDs) shown in Fig.~\ref{fig:MWL_SED} indicate a shift of both synchrotron and inverse-Compton (IC) peaks to higher energies during the 2015 observation campaign with respect to the archival data, accompanied by a significant variation of the X-ray and HE $\gamma$-ray spectral  indices. This behavior resembles the less extreme outburst seen in PMN~J2345--1555 \citep{Ghisellini2013}, and can be explained by a change in the emitting region location: within the broad line region (BLR) in the quiescent state to beyond the BLR during the outburst, where the external photon field is dominated by the optical--UV from the BLR or the IR thermal emission of a dusty torus,  respectively \citep[conventional framework for $\gamma$-loud FSRQ, e.g. ][]{GT09,Tavecchio2011}.

The consequences of this scenario are two-fold: (1) since the radiation energy density of the IR component is much lower than the one associated with the optical--UV photons from the BLR, the electron radiative cooling is less effective and the energy reachable by the acceleration process could be higher, accounting for the shift of the SED peak toward higher energies; (2) given the much lower energy of the external photons, absorption of $\gamma$ rays by pair production occurs only above several hundreds of GeV \citep[e.g., ][]{Protheroe1997}, enabling the detection of FSRQs at VHE. For an emission region well within the BLR, strong absorption features are expected for energies above tens of GeV \citep[see e.g.][]{Donea2003,Liu2006}, which are not observed in the 2015 MWL SEDs.

According to this framework, the proposed SED model for the 2015 observations assumes that the emission region is located at a distance $d>R_{\rm BLR}$ from the central compact object. Adopting the simple scaling proposed by \cite{GT09}, $R_{\rm BLR}$ depends only on the disk luminosity, $R_\mathrm{\rm BLR}=10^{17} (L_{disk}/10^{45})^{1/2}$ cm. The latter can be inferred from the luminosity of the optical broad lines, $L_{\rm disk}\simeq 2.0\times 10^{45}\, \mathrm{erg}\,\mathrm{s}^{-1}$ \citep{GT15}, resulting in $R_{\rm BLR}=1.4\times 10^{17}\ \mathrm{cm}$. In the same way, the size of the dusty torus can be inferred from a similar scaling law,  $R_{\rm IR}=3.5\times 10^{18}$ cm. The resulting emission is calculated using the code described in \cite{MT03}. The emission region is assumed to be spherical (in the source frame) with radius $R$, in motion with bulk Lorentz factor $\Gamma$ at angle $\theta_{\rm v}$ with respect to the line of sight. It contains a tangled magnetic field with intensity $B$ and a population of relativistic leptons described by a smoothed broken PWL energy distribution between Lorentz factors $\gamma_{\rm min}$ and $\gamma_{\rm max}$, with a break at $\gamma_{b}$, slopes $n_1$ and $n_2$ and normalization $K$ estimated at $\gamma$=1. The external photon field (dominated by the IR torus emission) is assumed to follow a black body spectrum with luminosity $L_{IR}=\xi L_{\rm disk}$ \citep[$\xi=0.6$, following][]{GT09} and temperature $T$, diluted within a region of radius $R_\mathrm{IR}$.

\begin{deluxetable*}{rrrrrrrrr}
\tabletypesize{\small}
\tablewidth{\textwidth}

\tablecaption{Upper limits at $95\,\%$ confidence level on the relative EBL opacity $\alpha$}
\tablenum{2}

\tablehead{\colhead{EBL model} & \colhead{Shape} & \colhead{$\mathrm{\alpha^{nominal}_{best}}$} & \colhead{$\mathrm{\alpha^{UL}_{w/syst}}$} & \multicolumn{4}{c}{Param. (best fit)} & \colhead{$p-\mathrm{value}$} \\ 
\colhead{} & \colhead{} & \colhead{} & \colhead{} & \colhead{$p_0$} & \colhead{$p_1$} & \colhead{$p_2$} & \colhead{$p_3$} & \colhead{} }

\startdata
PWL & No EBL & - & - & $-11.9$ & $-4.6$ & & & $<0.01$ \\
PWL & F08  & $1.09^{+0.36}_{-0.31}$ & $1.72$ & $-11.6$ & $-3.1$ & & & $0.50$ \\
PWL & D11  & $1.09^{+0.37}_{-0.32}$ & $1.73$ & $-11.5$ & $-3.1$ & & & $0.47$ \\
PWL & G12  & $0.99^{+0.33}_{-0.28}$ & $1.55$ & $-11.4$ & $-2.7$ & & & $0.51$ \\
PWL & S14 (max) & $1.09^{+0.37}_{-0.32}$ & $1.73$ & $-11.5$ & $-3.1$ & & & $0.47$ \\
PWL & S14 (min) & $2.20^{+0.70}_{-0.61}$ & $3.41$ & $-11.4$ & $-2.7$ & & & $0.54$ \\
\hline
LP & No EBL & - & - & $-11.9$ & $-4.7$ & $3.2$ & & $0.39$ \\
LP & F08  & $0.35^{+1.06}_{-1.58}$ & $1.69$ & $-11.8$ & $-4.2$ & $2.2$ & & $0.40$ \\
LP & D11  & $0.18^{+1.20}_{-1.42}$ & $1.68$ & $-11.8$ & $-4.4$ & $2.7$ & & $0.39$ \\
LP & G12  & $0.37^{+0.92}_{-1.63}$ & $1.53$ & $-11.7$ & $-3.9$ & $2.0$ & & $0.40$ \\
LP & S14 (max) & $0.18^{+1.20}_{-1.42}$ & $1.68$ & $-11.8$ & $-4.4$ & $2.7$ & & $0.39$ \\
LP & S14 (min) & $1.64^{+1.25}_{-3.56}$ & $3.40$ & $-11.5$ & $-3.2$ & $0.83$ & & $0.42$ \\
\hline
PWLsc & No EBL & - & - &  $-6.2$ & $1.4$ & $-0.41$ & $0.48$ & $0.27$ \\
PWLsc & F08  & $0.22^{+1.20}_{-3.21}$ & $1.70$ & $-7.4$ & $0.46$ & $-0.13$ & $0.47$ & $0.27$ \\
PWLsc & D11  & $0.15^{+1.23}_{-3.14}$ & $1.68$ & $-2.7$ & $2.7$  & $-1.9$ & $0.34$ & $0.27$ \\
PWLsc & G12  & $0.37^{+0.92}_{-3.36}$ & $1.54$ & $-1.4$ & $2.6$  & $-3.0$ & $0.27$ & $0.27$ \\
PWLsc & S14 (max) & $0.15^{+1.23}_{-3.14}$ & $1.68$ & $-2.7$ & $2.7$  & $-1.9$ & $0.34$ & $0.27$ \\   
PWLsc & S14 (min) & $1.75^{+1.15}_{-4.74}$ & $3.40$ & $-2.4$ & $0.39$ & $-5.8$ & $0.17$ & $0.29$
\enddata

\tablerefs{F08: \cite{Franceschini2008}, D11: \cite{Dominguez2011}, G12: \cite{Gilmore2012}, S14: \cite{Stecker2014}.}
\tablecomments{The normalization factor $10^{p_0}$ is given in units of $\mathrm{erg}\,\mathrm{cm}^{-2}\,\mathrm{s}^{-1}$.}

\label{tab:EBL}

\end{deluxetable*}

The model also includes $\gamma$-ray absorption within the IR radiation field of the torus. Assuming a temperature $T\approx 10^3\ \rm{K}$ for the IR torus, the maximum absorption is reached at $E=(m_{\rm e} c^2)^2/2.8 kT\simeq 1\ \mathrm{TeV}$ in the source frame, with an optical depth $\tau_{\gamma \gamma}\approx (\sigma_{\rm T}/5) (U_{IR}/h\nu_{IR}) R_{\rm IR}$ $\approx 250$. Given the large optical depth and the relatively broad spectrum of the target photons, the absorption is appreciable at few hundreds of GeV, i.e. $5\,\%$ at $200\,\mathrm{GeV}$ and $50\,\%$ at $300\,\mathrm{GeV}$ in the observer frame. Note also that an additional softening of the spectrum can be due to the fact that the emission in the VHE band is produced by scattering occurring in the Klein--Nishina (KN) regime \citep[e.g. ][]{Blumenthal1970, Zdziarski1993, Moderski2005}. 

To decrease the number of free parameters, we fix the bulk Lorentz and Doppler factor to $\Gamma=15$ and $\delta=20$, close to the average obtained for a large sample of $\gamma$-loud FSRQ \citep{GT15}. This implies a viewing angle of the jet $\theta_{\rm v}=2.7^{\rm o}$, and the aperture angle is fixed to $\theta_{\rm j}=0.1$ rad ($5.7^{\rm o}$).

We assume that the emission region is located beyond but not very far from the BLR, $d=5\times 10^{17}\,\mathrm{cm}$, implying $R=5\times 10^{16}\,\mathrm{cm}$. The low-energy slope $n_1$ is fixed to the standard value of 2. The remaining parameters are chosen to reproduce the synchrotron and IC components (see Table~\ref{tab:model}). To account for the different flux states, an evolution in both the electron distribution and the magnetic field is required. For comparison, the archival data (representation of the quiescent state) were modeled considering the emitting region partially within the BLR (standard framework) at $d=1.4\times 10^{17}$ cm, so that the $\gamma\gamma$ optical depth is small as indicated by the highest energy point of the 3FGL spectrum. The ratio between the peak luminosities (Compton Dominance, CD), are reported in Table~\ref{tab:model}. During the outburst, $\nu_{syn}$ lies more than an order of magnitude outside the FSRQ parameter space in the CD sequence proposed by \cite{Finke2013}, indicating a shift in the sequence during flares. The high degree of optical polarization suggests that the emission may come from a compressed region in the jet like an internal shock, which is also an ideal site for electron acceleration/injection.


\section{Extragalactic background light constraints}

VHE $\gamma$ rays from distant blazars can interact with the optical--UV photons from the EBL via pair production \citep{Gould1967,Stecker1992}, resulting in an attenuation of the intrinsic VHE spectrum. The EBL imprint in the $\gamma$-ray spectra from distant blazars can be used to constrain the EBL density.

\begin{figure*}
  \centering
  \includegraphics[width=0.62\textwidth]{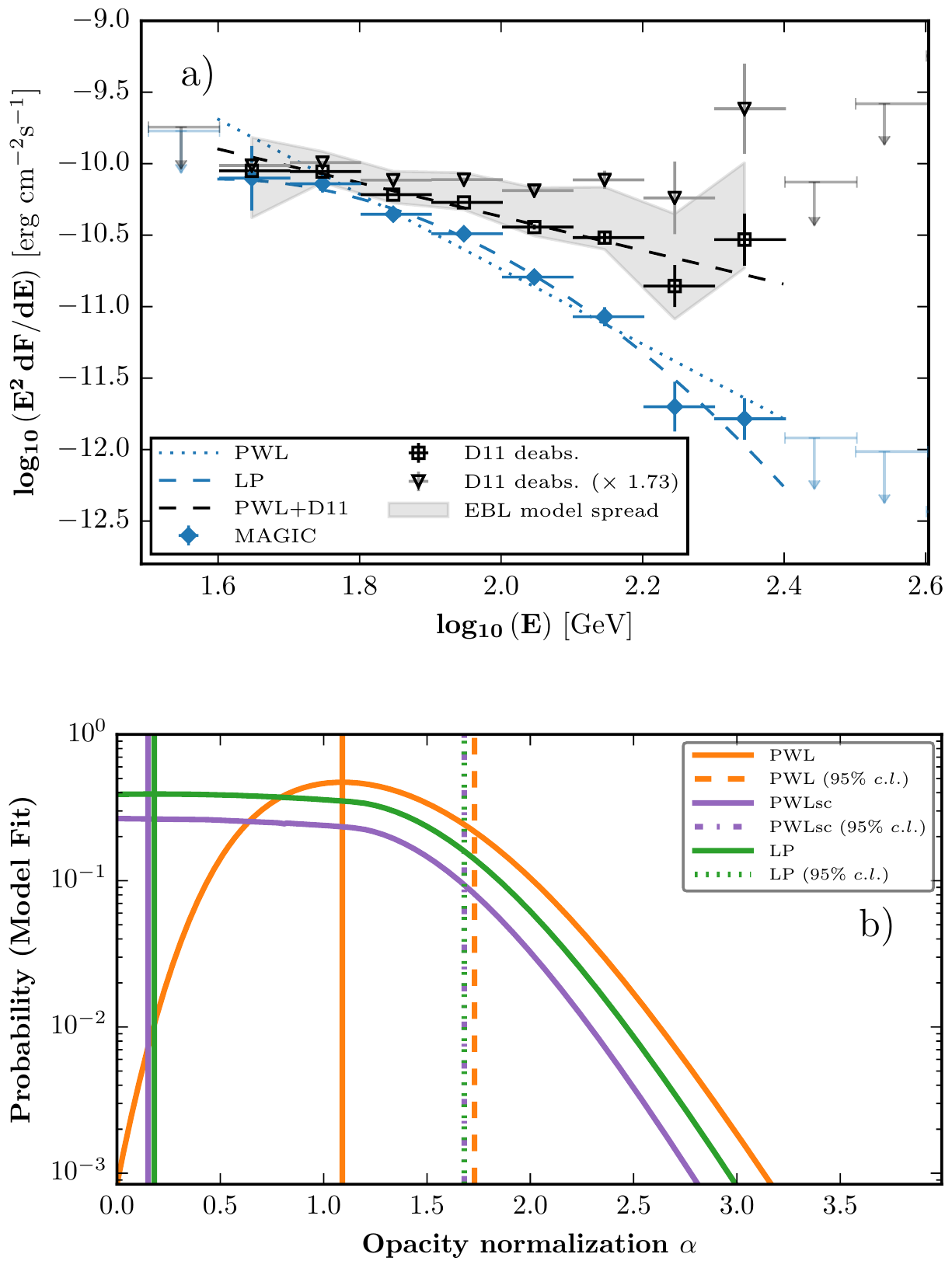}
  \caption{(a) Observed (blue diamonds) and EBL-corrected SED using \cite{Dominguez2011} (black squares) for period B. The dotted and dashed lines show the best-fitting PWL, respectively. The gray shaded area accounts for the uncertainties derived by the use of different EBL models \citep{Franceschini2008,Dominguez2011,Gilmore2012}. (b) The probability of fit as a function of EBL relative opacity \citep[D11]{Dominguez2011}.  Only period \emph{B} was considered (without upper limits). The best fit is marked with solid vertical lines and $95\,\%$ confidence level upper limits with dashed vertical lines.}
  \label{fig:EBL_SED}
\end{figure*}

Measurements of the EBL absorption can be derived under some assumptions on the intrinsic spectrum of the source \citep[see e.g. ][]{Ackermann2012_EBL,Abramowski2013_EBL}. With a redshift of $z=0.940$ and a strong detection in the VHE band, PKS~1441+25 allows us to probe EBL models at a distance never explored before in this energy regime with ground-based gamma-ray instruments. However, KN effects together with an expected intrinsic $\gamma$-ray absorption in the VHE band (see Sec.~\ref{BroadbandSED}), can mimic the effect of EBL absorption, making it difficult to disentangle the two effects.

A LRT was used to compare a null hypothesis (no EBL absorption) with respect to the hypothesis of EBL absorption with a scaled opacity $\alpha\, \tau(z,E)$ as in \cite{Abdo2010_EBL}. Predicted opacities from \cite{Dominguez2011} ($\tau_{D11}$), \cite{Franceschini2008} ($\tau_{F08}$), \cite{Gilmore2012} ($\tau_{G12}$) and \cite{Stecker2014} ($\tau_{S14}$) are considered, while $\alpha$ is a free scaling parameter. Different intrinsic spectral shapes were assumed: PWL $dF/dE = 10^{p_0} (E/E_0)^{p_1}$, LP $dF/dE = 10^{p_0} (E/E_0)^{p_1 - p_2 \log_{10}{E/E_0}}$ and PWLsc $dF/dE = 10^{p_0} (E/E_0)^{p_1} \exp{[(E/10^{p_2})^{p_3})]}$ where $E$ is measured in $\mathrm{GeV}$ and $E_0 = 100\, \mathrm{GeV}$. The limits are reported in Table \ref{tab:EBL} and an example is given in Fig.~\ref{fig:EBL_SED}. A possible overall systematic error of $\pm 15\,\%$ in the absolute energy scale of the instrument is considered. Under the assumption that no curvature is present in the intrinsic VHE spectrum, the measured spectrum is compatible with the present generation of EBL models. 

The $95\,\%$ confidence level limit obtained in this work for \cite{Franceschini2008} is compatible with the one found in \cite{Ackermann2012_EBL} for $0.5\leq z\leq 1.6$, $\alpha=1.3\pm0.4$, which is obtained from observations with a wide range of redshift values while our UL is calculated for a precise redshift value.

The estimated scaling on the optical depth can be translated into EBL density constraints as shown in \cite{Dominguez2011} and \cite{Abramowski2013_EBL}. The observed VHE spectrum allow us to constrain the EBL density between $0.21$ and $1.13\,{\rm \mu m}$, where the optical depth with respect to the nominal value of \cite{Dominguez2011}, $\alpha_{D11}<1.73$, implies in the local Universe $\lambda f_{\lambda=0.5\mu m}< 8.7\,\rm{nW} \rm{cm}^{-2} \rm{sr}^{-1}$.

\section{Conclusions}

MAGIC has detected for the first time VHE emission from the $z=0.940$ blazar PKS~1441+25 during a MWL outburst in 2015 April. PKS~1441+25 is, together with QSO~B0218+357, the most distant VHE source detected so far. This allow us to study VHE blazars when the Universe was only half of its current age.

The evolution of the MWL SED is studied in the framework of an external Compton emission model. The absence of intrinsic absorption features in the HE and the VHE regime constrains the localization of the emitting region to be just outside of the BLR during this period of high activity, while it is expected to be partially compatible with the BLR during the period of low activity. The SED evolution reveals changes in the electron distribution and the magnetic field.

For the first time, the VHE measurements are used to indirectly probe the EBL at redshifts out to $\mathrm{z}\sim 1$ with ground-based gamma-ray instruments. Although an internal cutoff cannot be excluded, the measured VHE spectrum is consistent with a steepening due to attenuation caused by the EBL. Employing state-of-the-art EBL models, upper limits to the EBL density are derived. The upper limits on the opacity calculated under the assumption of an intrinsic spectrum compatible with a PWL function for different EBL models result in $\tau(z,E)<1.73\,\tau_{D11}$, $\tau(z,E)<1.72\,\tau_{F08}$, $\tau(z,E)<1.55\,\tau_{G12}$, $\tau(z,E)<1.73\,\tau_{S12_{max}}$ and $\tau(z,E)<3.41\,\tau_{S12_{min}}$ for EBL models from \cite{Dominguez2011}, \cite{Franceschini2008}, \cite{Gilmore2012} and maximum and minimum from \cite{Stecker2014}, respectively.

\section{Acknowledgments}

We would like to thank the Instituto de Astrof\'{\i}sica de Canarias for the excellent working conditions at the Observatorio del Roque de los Muchachos in La Palma. The financial support of the German BMBF and MPG, the Italian INFN and INAF, the Swiss National Fund SNF, the ERDF under the Spanish MINECO (FPA2012-39502) and MECD (FPU13/00618), and the Japanese JSPS and MEXT is gratefully acknowledged. This work was also supported by the Centro de Excelencia Severo Ochoa SEV-2012-0234, CPAN CSD2007-00042, and MultiDark CSD2009-00064 projects of the Spanish Consolider-Ingenio 2010 programme, by grant 268740 of the Academy of Finland, by the Croatian Science Foundation (HrZZ) Project 09/176 and the University of Rijeka Project 13.12.1.3.02, by the DFG Collaborative Research Centers SFB823/C4 and SFB876/C3, and by the Polish MNiSzW grant 745/N-HESS-MAGIC/2010/0.

The {\it Fermi}-LAT Collaboration acknowledges support for LAT development, operation and data analysis from NASA and DOE (United States), CEA/Irfu and IN2P3/CNRS (France), ASI and INFN (Italy), MEXT, KEK, and JAXA (Japan), and the K.A. Wallenberg Foundation, the Swedish Research Council and the National Space Board (Sweden). Science analysis support in the operations phase from INAF (Italy) and CNES (France) is also gratefully acknowledged.

We thank the {\it Swift} team duty scientists and science planners. L.P. acknowledges the PRIN-INAF 2014 financial support.

\end{document}